\documentclass [12pt]{article}
\RequirePackage[english,russian]{babel}
\RequirePackage[cp1251]{inputenc}
\usepackage{amsmath,amssymb}

\begin{document}
\sloppy
\begin{center}

F.A. Gareev,  I.E. Zhidkova\\
  Cooperative Mechanisms of Low Energy  Nuclear Reactions Using Superlow Energy
External Fields\\
Joint Institute for Nuclear Research, Dubna, Russia\\
 e-mail:gareev@thsun1.jinr.ru\\
\end{center}

\section{Introduction}
 The  review of possible stimulation mechanisms of LENR (low energy nuclear reaction) is presented in \cite{2}.
We have concluded that transmutation of nuclei at low energies and
excess heat are possible in the framework of the known fundamental
physical laws – the universal resonance synchronization principle
\cite{1}, and different enhancement mechanisms of reaction rates
based on it are responsible for these processes \cite{2}. The
superlow energy of external fields, the excitation and ionization
of atoms may play the role of a trigger for LENR. Superlow energy
of external fields may stimulate LENR \cite{3}. We bring  strong
arguments that the cooperative mechanism is responsible for
explanation of how the electron volt world can  influence  the
nuclear mega electron volt world \cite{3}. Nuclear physicists are
absolutely sure that this is cannot happen. Investigation of this
phenomenon requires the knowledge of different branches of
science: nuclear and atomic physics, chemistry and
electrochemistry, condensed matter and solid state physics,...
 The puzzle of poor reproducibility of experimental data is the fact that LENR occurs in open systems and
it is extremely sensitive to parameters of external fields and
systems. The classical reproducibility principle should be
reconsidered for LENR experiments. Poor reproducibility and
unexplained results do not mean that the experiment is wrong.  Our
main conclusion is:
 LENR may be understood in terms of the known fundamental laws without any violation of  basic physics.
The fundamental laws of physics should be  the same in micro- and
macrosystems.

Let us start with the description of  the hydrogen atom structure
in different models.

\subsection{The Hydrogen  Atom}
We will  describe very shortly  the structure of a hydrogen atom
using  standard basic physics that is well established, both
theoretically and experemintally in micro- and  macrosystems.

\subsection{The Bohr Model}

At the end of the 19th century it was established that the
radiation from hydrogen was emitted at specific quantized
frequencies. Niels Bohr developed the model to explain this
radiation using four postulates:

1. An electron in an atom moves in a circular orbit about the nucleus under the influence of the
Coulomb attraction between the electron and the nucleus, obeying the laws of classical mechanics.

2. Instead of the infinity of orbits which would be possible in
classical mechanics, it is only possible for an electron to move
in an orbit for which its orbital angular momentum $L$ is
integral multiple of $\hbar$:
                                    $$ L=n\hbar,\;\;n=1,2,3,…\eqno(1) $$

3. Despite the fact that it is constantly accelerating, an
electron moving in  such an allowed orbit does not radiate
electromagnetic energy. Thus, its total energy $E$ remains
constant.

4. Electromagnetic radiation is emitted if an electron, initially
moving in an orbit of total energy $E_{i}$, discontinuously
changes its motion so that it moves in an orbit of total energy
$E_{f}$. The frequency $\nu$ of the emitted radiation is equal to
the quantity

$$\nu_{if}=\frac{E_{i}-E_{f}}{h},\eqno(2)$$

where $h$ is Planck's constant.
 The electron is held in a stable
circular orbit around a nucleus. The Coulomb force is equal to the
centripetal force, according to Newton's second law

$$\frac{e^{2}}{r^{2}}=\frac{mv^{2}}{r},\eqno(3)$$
where $r$ is is the radius of the electron orbit, and $v$ is the
electron speed. The force is central; hence from the quantization
condition (1) we have

$$L=\mid \vec{r}*\vec{p}\mid=mvr=n\hbar.\eqno(4)$$

After solving equations (3) and (4) we have

$$v=\frac{e^{2}}{n\hbar},\;r=\frac{n^{2}\hbar^{2}}{me^{2}}=n^{2}a_{0}.\eqno(5)$$
Following equation (3) the kinetic energy is equal to

$$E_{k}=\frac{1}{2}mv^{2}=\frac{e^{2}}{2r},\eqno(6)$$
and hence the total energy is

$$E=E_{k}+V=\frac{e^{2}}{2r}-\frac{e^{2}}{r}=-\frac{e^{2}}{2r}.\eqno(7)$$

Having $r$ from equation (5) one can write the expression for the
energy levels for  hydrogen atoms

$$E=-\frac{me^{4}}{2\hbar^{2}n^{2}};\eqno(8)$$
the same results were further obtained  by quantum mechanics.

Using the angular momentum quantization condition $L=pr=nh/2\pi$
and Louis de Broglie's relationship $p=h/\lambda$  between
momentum and wavelength one can get

$$2\pi r=n\lambda. \eqno(9)$$

{\sl $\otimes$ It means that the circular Bohr orbit is an
integral number of the de Broglie wavelengths. The Bohr model is
actually only accurate for a one-electron system. }

\subsection{ The Hydrogen Atom in Classical Mechanics}

Is it possible to understand some properties of a hydrogen atom
from classical mechanics ? The Hamiltonian for a hydrogen atom is

$$H=\frac{m_p \dot{\vec{r_p}}\;^2}{2} +
\frac{m_e \dot{\vec{r_e}}\;^2}{2} - \frac{e^2}{ \mid \vec{r}_p -
\vec{r}_e \mid }.\eqno(10)$$

All notation is standard. The definition of the center of mass is

$$m_{p}\vec{r}_{p}+m_{e}\vec{r}_{e}=0, \eqno(11) $$

and the relative distance between electron and proton is

$$\vec{r}=\vec{r}_p- \vec{r}_e. \eqno(12)$$

Equations (10)-(12) lead to the results:

$$\vec{r}_{p}=\frac{m_{e}}{m_{p}+m_{e}}\vec{r},\;\vec{r}_{e}=
-\frac{m_{p}}{m_{p}+m_{e}}\vec{r}, \eqno(13)$$

$$H=\frac{\mu \dot{{\vec
r}}\;^2}{2}-\frac{e^{2}}{r},\eqno(14)$$

where
$$\mu=\frac{m_{p}m_{e}}{m_{p}+m_{e}}.\eqno(15)$$

The Hamiltonian (14) coincides with the Hamiltonian for the
fictitious material point with reduced mass $\mu$ moving in the
external field $-e^{2}/r$. If we known the trajectory of this
fictitious particle $\vec{r}=\vec{r}(t)$ then we can reconstruct
the trajectories of electron and proton using equations (13)

$$\vec{r}_{p}(t)=\frac{m_{e}}{m_{p}+m_{e}}\vec{r}(t),\;\;\;
\vec{r}_{e}(t)=-\frac{m_{p}}{m_{p}+m_{e}}\vec{r}(t).\eqno(16)$$

It is evident from (16) that the proton and electron  move in the
opposite directions synchronously. So the motions of proton,
electron and their relative motion occur with  equal frequency

$$\omega_{p}=\omega_{e}=\omega_{\mu},\eqno(17)$$
over the closed trajectories scaling by the ratio

$$ \frac{v_{e}}{v_{p}}=\frac{m_{p}}{m_{e}},\;\frac{v_{e}}{v_{\mu}}=\frac{m_{\mu}}{m_{e}},\;
\frac{v_{\mu}}{v_{p}}=\frac{m_{p}}{m_{\mu}}.\eqno(18)$$

I.A. Schelaev \cite{SCH04} proved that the frequency spectrum of
any motion on ellipse contains only one harmonic.

We can get from (16) that
$$\vec{P}_{p}=\vec{P},\;\vec{P}_{e}=-\vec{P}, \eqno(18a)$$

where -- $\vec{P}_{i}=m_{i}\vec{\dot{r}}_{i}$. All three impulses
are equal to each other in absolute value, which means the
equality of
$$\lambda_{D}(p)=\lambda_{D}(e)=\lambda_{D}(\mu)=h/P.\eqno(19)$$

Conclusion:\\
{\sl $\otimes \;\;\;$ Therefore, the motions of proton and
electron and their relative motion occur with the same FREQUENCY,
IMPULSE (linear momentum) and the de Broglie WAVELENGTH. All
motions are synchronized and self-sustained. Therefore, the whole
system -hydrogen atom is nondecomposable to the independent
motions of proton and electron despite the fact that the kinetic
energy ratio of electron to proton  is small:

$$ \frac{E_{k}(e)}{E_{k}(p)}=4.46*10^{-4}.$$
It means that the nuclear and the corresponding atomic processes
must be considered as a unified  entirely determined whole
process.

For example, V.F. Weisskopf \cite{6} came to the conclusion that
the maximum height $H$ of mountains in terms of the Bohr radius
$a$ is equal to

$$\frac{H}{a}=2.6*10^{14},$$
and water wave lengths $\lambda$ on the surface of a lake in terms
of the Bohr radius is equal to}

$$\frac{\lambda}{a}\approx 2\pi*10^{7}.$$

Let us introduce the quantity $f=rv$ which is the invariant of
motion, according to the second Keplers law, then

$$ \mu v=\frac{\mu vr}{r}=\frac{\mu f}{r},\eqno(20)$$
and we can rewrite  equation (14) in the following way:

$$ H=\frac{\mu f^{2}}{2r^{2}}-\frac{e^{2}}{r}. \eqno(21)$$

We can obtain the minimal value of (21) by taking its first
derivative over $r$ and setting it equal to zero. The minimal
value occurs at

$$ r_{0}=\frac{\mu f^{2}}{e^{2}}, \eqno(22)$$
and the result is

$$ H_{min}=E_{min}=-\frac{e^{4}}{2 \mu f^{2}}.\eqno(23)$$

The values of invariant of motion  $\mu f$ (in MeV*s) can be
calculate from (23) if we require the equality of  $E_{min}$ to
the energy of the ground  state of a hydrogen atom

$$ \mu f= \mu
vr=6.582118*10^{-22}=\hbar,\;\eqno(24)$$

Conclusion:\\
{\sl $\otimes \;\;\;$ The Bohr quantization conditions were
introduced as a hypothesis. We obtain these conditions from a
classical Hamiltonian requiring its minimality. It is necessary to
strongly stress that no assumption was formulated about
trajectories of proton and electron. We reproduced exactly the
Bohr result and modern quantum theory. The Plank constant $\hbar$
is the Erenfest adiabatic invariant for a hydrogen atom: $\mu vr =
\hbar$.}

Let us briefly review our steps:

$\bullet$ We used a well established interaction between proton
and electron.

$\bullet$ We used a fundamental fact that the total energy=kinetic
energy+potential energy.

$\bullet$ We used the second Kepler law.

$\bullet$ We used usual calculus to determine the minimum values
of $H$.

$\bullet$ We required the equality of  $E_{min}$ to the energy of
the ground state of hydrogen atom.

 Classical Hamiltonian + classical interaction between proton and
electron + classical second Kepler law + standard variational
calculus -- these well established steps in macrophysics reproduce
exactly  results of the Bohr model and modern quantum theory
(Schrodinger equation) -- results of microphysics. We have not
done anything spectacular or appealed to any revolutionary and
breakthrough physics.

Using the Newton equation with well established interactions M.
Gryzinski \cite{GRY04} proved that atoms have the quasi-crystal
structure with definite angles: $90^{\circ}$, $109^{\circ}$ and
$120^{\circ}$, which are the well-known angles in crystallography.

\section{Nuclei and Atoms as Open Systems}

1) LENR may be understood in terms of the known fundamental laws
without any violation of the basic physics. The fundamental laws
of physics should be  the same in micro- and macrosystems.

2)Weak and electromagnetic interactions may show a strong
influence of the surrounding conditions on the nuclear
processes.\\

3)The conservation laws are valid for  closed systems. Therefore,
the failure of parity in weak interactions means that the
corresponding systems are  open systems. Periodic variations (24
hours, 27, and 365 days in  beta-decay rates indicate that the
failure of parity in weak interactions has a cosmophysical origin.
Modern quantum theory is the theory for closed systems. Therefore,
it should be reformulated for open systems. The closed systems are
idealization of nature,  they do not exist in reality. \\

4)The universal resonance synchronization principle  is a key
issue to make a bridge between various scales of interactions and
it is responsible for self-organization of hierarchical  systems
independent of substance, fields,  and interactions. We give some
arguments in favor of the mechanism – ORDER BASED on ORDER,
declared  by Schrodinger in \cite{4}, a fundamental problem of
contemporary science.\\

5)The universal resonance synchronization principle became a
fruitful interdisciplinary science of general laws of
self-organized processes in different branches of physics because
it is the consequence of the energy conservation law and resonance
character of any interaction between wave systems. We have proved
the homology of  atom, molecule and crystal structures including
living cells. Distances of these systems are commensurable  with
the de Broglie wave length  of an electron in the ground state of
a hydrogen atom,  it plays the role of the standard distance, for
comparison. \\

6)First of all, the structure of a hydrogen atom should be
established. Proton and electron in a hydrogen atom move with the
same frequency that creates attractive forces between them, their
motions are synchronized. A hydrogen atom represents the radiating
and accepting  antennas (dipole) interchanging  energies with the
surrounding  substance. The sum of radiate and absorb energy flows
by electron and  proton in a stable orbit is equal to zero
\cite{5} – the secret of success of  the Bohr model (nonradiation
of  the electron on stable orbit). “The greatness of mountains,
the finger sized drop, the shiver of a lake, and the smallness of
an atom are all related by simple laws of nature” – Victor F. Weisskopf  \cite{6}.\\

7)These flows created  standing waves due to the resonance
synchronization principle. A constant energy exchange with
substances (with universes) create stable auto-oscillation systems
in which the frequencies of  external fields and all subsystems
are commensurable. The relict radiation (the relict isotropic
standing waves at T=2.725 K – the Cosmic Microwave Background
Radiation (CMBR))  and   many isotropic standing waves in cosmic
medium \cite{7} should be results of  self-organization of the
stable hydrogen atoms, according to the universal resonance
synchronization principle that is a consequence of  the
fundamental energy conservation law. One of the fundamental
predictions of the Hot Big Bang theory for the creation of the
Universe is CMBR.\\

8)The cosmic isotropic standing waves (many of them are not
discovered yet) should play the role of a conductor responsible
for stability of elementary particles, nuclei, atoms,…, galaxies
ranging in size more than 55 orders of magnitude.\\

9)The phase velocity of standing microwaves can be extremely high;
therefore, all objects of the Universe should  get information
from each other almost immediately using phase velocity.

The aim of this paper is to discuss the possibility of inducing
and controlling nuclear reactions at low temperatures and
pressures by using different low energy external fields and
various physical and chemical processes. The main question is the
following: is it possible to enhance LENR rates by using  low and
extremely low energy external fields? The review of possible
stimulation mechanisms is presented in \cite{2,5}. We will discuss
new  possibilities to enhance LENR rates in condensed matter.

\section{LENR in Condensed Matter}

The modern understanding of the decay of the neutron is

$$n \rightarrow p+e^{-}+\overline{\nu}_{e}.\eqno(25)$$
The energetics  of the decay can be analyzed using the concept of
binding energy and the masses of  particles by their rest mass
energies. The energy balance from neutron decay can be calculated
from the particle masses. The rest mass difference ( $0.7823
MeV/c^{2}$) between neutron and (proton+electron) is converted to
the kinetic energy of proton, electron, and neutrino. The neutron
is about $0.2\%$ more massive than a proton, an mass difference is
1.29 $MeV$. A free neutron will decay with a half-life of about
10.3 minutes. Neutron in a nucleus will decay if a more stable
nucleus results; otherwise neutron in a nucleus will be stable. A
half-life of neutron in nuclei changes dramatically and depends on
the isotopes.

The capture of electrons by protons

$$p+e^{-}\rightarrow n+\nu_{e},\eqno(26)$$
but for free protons and electrons this reaction has never been
observed which is the case in nuclear+ atomic physics. The capture
of electrons by protons in a nucleus will occurs if a more stable
nucleus results. \\

\subsection{Cooperative Processes}

The processes (25) and (26) in LENR  are going with individual
nucleons and electrons. In these cases the rest mass difference is
equal to $0.7823 MeV/c^{2}$. In the case of neutron decay the
corresponding energy ($Q=0.7823$ MeV) converted to kinetic
energies of proton, electron, and antineutrino. In the case of the
capture of electrons by protons the quantity  $Q=0.7823$ MeV is a
threshold electron kinetic energy under which the process (26) is
forbidden for free proton and electron.

We have formulated the following postulate:\\
$\otimes$ {\sl The processes (25) and (26) in LENR  are going  in
the whole system: cooperative processes including all nucleons in
nuclei and electrons in atoms, in condensed matter. In these cases
a threshold energy $Q$ can be drastically decreased by internal
energy of the whole system or even more -- the electron capture by
proton can be accompanied by emission of internal binding energy -
main source of excess heat phenomenon in LENR. }

The processes (25) and (26) are weak processes. A weak interaction
which is responsible for electron capture and other forms of beta
decay  is of a very short range. So the rate of electron capture
and emission (internal conversion) is proportional to the density
of electrons in nuclei. It means that we can manage the
electron-capture (emission) rate by the change of the total
electron density in the
nuclei using different low energy external fields. These fields \\
can play a role of triggers for extracting internal energy of the
whole system or subsystems, changing quantum numbers of the
initial states in such a way that forbidden transitions become
allowed ones. The distances between proton and electron in atoms
are  of the order $10^{-6}--10^{-5}$ cm and any external field
decreasing these distances even for a small value can increase the
process (26) in nuclei in an exponential way. Therefore, the
influence of an external electron flux (discharge in condensed
matter: breakdown, spark and ark) on the velocity processes (25)
and (26) can be of great importance.

The role of external electrons is the same as the catalytic role
of neutrons in the case of the chain fission reactions in nuclei
-- neutrons bring to nuclei binding energies (about 8 MeV) which
enhance the fission rates by about 30 orders.

\section{Predicted Effects and Experimentum  Cruices}

Postulated enhancement mechanism of LENR by external fields can be
verified by the Exprimentum Cruices. We \cite{5} predicted that
natural geo-transmutation in the atmosphere and earth occur in the
the regions of a strong change in geo-, bio-, acoustic-,...  and
electromagnetic fields.

Various electrodynamic processes at thunderstorms are responsible
for different phenomena: electromagnetic pulses, $\gamma$-rays,
electron fluxes, neutron fluxes, and radioactive nuclei fluxes.

\subsection{Neutron Production by Thunderstorms}

 The authors of  \cite{BRA4} concluded that a  neutron burst is
associated with lighting. The total number of neutrons produced by
one typical lightning discharge was estimated as $2.5*10^{10}$.

\subsection{Production of Radiocarbon and Failing of Radiocarbon
Dating}

The radiocarbon dating is based on the decay rate of radioactive
isotope $^{14}C$ which is believed to be constant irrespective of
the physical and chemical conditions. The half-life of radiocarbon
$^{14}C$ is 5730 years. A method for historical chronometry was
developed assuming that the decay ratio of $^{14}C$ and its
formation are constant in time. It was postulated that $^{14}C$ is
formed only by the cosmic ray neutrons

$$^{14}N(n,p)^{14}C. \;\eqno(27)$$
Radiocarbon dating is widely used in archeology, geology,
antiquities,... There are over 130 radiocarbon dating
laboratories. The radiocarbon method of dating was developed by
Willard F. Libby who was awarded  the Nobel prize in Chemistry for
1960.

The radiocarbon method does not take into account the following
facts which have been established recently:

$\otimes$ The neutron production by thunderstorms \cite{BRA4}

$\otimes$ The Production of radiocarbon by lighting bolts
\cite{LIB73}.

 Let us consider the reaction

$$^{14}_{7}N+e^{-}\rightarrow ^{14}_{6}C+\nu_{e},\eqno(27a)$$
$T_{k}(e)$=156.41 keV is the threshold energy which should by
compared with 782.3 keV for process (26). Production of
radiocarbon by lighting bolts was established in \cite{LIB73}.
Unfortunately, this means FAILING of RADIOCARBON DATING.

\subsection{Production Radiophosphorus by Thunderstorms}

The life-times of $^{32}_{15}P$ and $^{33}{15}P$ are equal to
14.36 and 25.34 days, respectively. They were found in rain-water
after thunderstorms \cite{SEL70}. Production of the
radiophosphorus by thunderstorms can be understood in the
following way:

$$^{32}_{16}S+e^{-}\rightarrow ^{32}_{15}P+\nu_{e},\eqno(28)$$
$$^{33}_{16}S+e^{-}\rightarrow ^{33}_{15}P+\nu_{e},\eqno(29)$$
thresholds of these processes are equal to 1.710 and 0.240 MeV,
respectively.  The precipitation of MeV electrons from the inner
radiation belt \cite{INAN} and enhancement of the processes by
lightning are possible.

\subsection{LENR Stimulated by Condensed Matter Discharge}

Let us consider the condensed matter discharge (breakdown, spark
and arc) using the
different electrode.  There are the following processes:\\

1. The electrode is $Ni$. Orbital or external electron capture
$$^{58}_{28}Ni(68.27\%)+e^{-}\rightarrow ^{58}_{27}Co(70.78\;
days)+\nu_{e}, \eqno(30)$$

The threshold $Q_{1}=0.37766\;keV$ of this reaction on $Ni$ should
be compared with the threshold $Q_{2}=0.7823$ energy for electron
capture by free protons: $Q_{2}/Q_{1}\approx 2$. The velocity of
orbital electron capture can be enhanced by the discharge.

2.Orbital or external electron capture
$$^{58}_{27}Co(70.78\; days)+e^{-}\rightarrow ^{58}_{26}Fe(0.28\%)+\nu_{e},\eqno(31)$$
with emission of energy $Q_{2}=2.30408$ MeV.

3. Double orbital or external electron capture
$$^{58}_{28}Ni(68.27\%)+2e^{-} \rightarrow ^{58}_{26}Fe(0.28\%)+2\nu_{e},\eqno(32)$$
with emission of energy $Q_{3}=1.92642$ mostly by neutrinos.

The proposed cooperative mechanism of LENR in this case can be
proved in an extremely simple way: presence of radioactive
$^{58}_{27}Co$ and enriched isotope of $^{58}_{26}Fe$.

$\otimes$ {\sl This mechanism can give  possibilities to get a way
of controlling  the necessary isotopes and excess heat.}

\subsection{Neutrinoless Double Beta Decay}

As we known \cite{2}, the physical roles of electron and neutrino
for LENR in condensed matter has not been investigated in detail
up to now despite the fact that  weak processes in nuclei are well
understood. The double beta decay is the rarest spontaneous
nuclear transition,in which the nuclear charge changes by two
units while the mass number remains the same. Such a case can
occur for the isobaric triplet $A(Z,N)$, $A(Z\pm 1,N\mp 1)$,
$A(Z\pm 2,N\mp 2)$, in which the middle isobar has a greater rest
mass than the extreme ones, and the extremes are the nuclei with
the even $Z$ and $N$. The usual beta-decay transferring a given
nucleus into another via an intermediate nucleus is energetically
forbidden.

The double beta decay in nuclei  can proceed in different modes \cite{KLA95}: \\

$\otimes$The two neutrinos decay mode $2\nu\beta\beta$

$$A(Z,N)\rightarrow A(Z+2,N-2)+2e^{-}+2\overline{\nu}_{e},\eqno(33)$$
which is allowed by the Standard Model of particle physics. The
total kinetic energy of two emitted electrons present  continuous
spectra up to $E_{max}$.

$\otimes$The neutrinoless mode $0\nu\beta\beta$

$$A(Z,N)\rightarrow A(Z+2,N-2)+2e^{-},\eqno(34)$$
which requires violation of a lepton number. The total kinetic
energy of two emitted electrons is equal to $E_{max}$.

Two neutrinos in the mode $2\nu\beta\beta$ carry out almost all
emitted energies.  A fundamental question is: Does the
neutrinoless double beta decay exist or not (for the review of the
history see  \cite{KLA95,KLA05})?. The emerged energies in the
neutrinoless $0\nu\beta\beta$ mode are easily  detected for
practical use but these are the rarest spontaneous nuclear
transitions ($T\approx 10^{18}-10^{30}$ years). Is it possible to
enhance the decay rate?

Above and in \cite{1,2,3} we have discussed the cooperative and
resonance synchronization enhancement mechanisms of LENR. Some of
the low energy external fields can be used as triggers for
starting and enhancing of exothermic LENR. It is natural to expect
that in the case of beta-decay (capture)  the external electron
flux with high density, or the laser of high intensity, or any
suitable external fields should play this a role. Any external
field shortening distances between protons in nuclei and electrons
in atoms should enhance beta-decay (capture) or double-beta decay
(capture).

There are a great number of experiments in Japan, Italy, Russia,
US, India, China, Israel, and Canada in which cold transmutations
and excess energy were measured (see http://www.lenr-canr.org).
Indeed the existence of LENR is now well established but the
proposed about 150 theoretical models for interpretation of
experimental data are not accepted (A. Takahashi, ICCF12).

It is  very popular to use $Ni$, $Pd$, $Pt$ and $W$ as electrodes
in the condensed matter discharge (breakdown, spark, arc, and
explosion) experiments. Let us consider the case $Pd$ electrodes.
The difference of the rest mass of
$$m(^{110}_{46}Pd)-m(^{110}_{48}Cd)=1.9989\; MeV/c^{2};$$
therefore, the external field can open the channel
$^{110}_{46}Pd\rightarrow ^{110}_{48}Cd$ with $Q=1.9989$ MeV. In
the cases $Ni$, $Pt$, and $W$ we have

$$m(^{58}_{28}Ni)-m(^{58}_{26}Fe)=1.92642\;MeV/c^{2},$$
$$m(^{186}_{74}W)-m(^{186}_{76}Os)=0.47302\;MeV/c^{2},$$
$$m(^{198}_{78}Pt)-m(^{198}_{80}Hg)=1.05285\;MeV/c^{2}.$$

The proposed cooperative mechanism of LENR in these cases can be
proved in an extremely simple way: presence of  enriched isotopes
of $^{58}_{26}Fe$, $^{110}_{48}Cd$, $^{186}_{76}Os$, and
$^{198}_{80}Hg$ for the indicated above electrodes.

The experimental data \cite{SAV96,KRI03,KUZ03,KRY02,BAL03} seem to
confirmed such expectations.

Therefore,  expensive and time consume  double beta decay
experiments can be performed in extremely  cheap and short-time
experiments by using suitable external fields. This new direction
of research can give answers for fundamental problems of modern
physics (the lepton number conservation, type of neutrino,
neutrino mass spectrum,... ), it can open production of new
elements (utilization of radioactive waste) and excess heat
without of ecological problem.

 A careful analysis of the double
beta decay shows that the $2e^{-}$ cluster can be responsible for
the double beta decay. The difference between the rest mass
$^{130}_{56}Ba$ and $^{130}_{52}Te$, which is equal to 92.55 keV,
indicates the possibilities to capture the $4e^{-}$ cluster by
$^{130}_{56}Ba$. It is a full analogy with the Iwamura reactions
\cite{IWA11}.

The lack of financial support and the ignorance from the whole
physical society of LENR lead to the catastrophes. The mechanism
of shortening the runaway of the reactor at the Chernobyl Nuclear
Power Plant and catastrophes induced by the HAARP (High Frequency
Active Auroral Research Program) program is based on our
postulated cooperative resonance synchronization mechanism. The
same mechanism should be responsible for the ITER (The
International Thermonuclear Experimental  Reactor) explosion in
future.

\section{Conclusion}

We proposed a new mechanism of LENR: cooperative processes in
whole system - nuclei+atoms+condensed matter can occur at smaller
threshold energies then corresponding ones on free constituents.
The cooperative processes can be induced and enhanced by low
energy external fields. The excess heat is the emission of
internal energy and transmutations at LENR are the result of
redistribution inner energy of whole system.

\end{document}